\documentclass[11pt]{article}
\newcommand{\nc}{\newcommand}
\nc{\bib}{\bibitem}
\nc{\al}{\alpha}
\nc{\g}{\gamma}
\nc{\G}{\Gamma}
\nc{\D}{\Delta}
\nc{\eps}{\epsilon}
\nc{\la}{\lambda}
\nc{\La}{\Lambda}
\nc{\var}{\varphi}
\nc{\pa}{\partial}
\nc{\nn}{\nonumber \\ }
\nc{\hf}{\frac{1}{2}}         
\nc{\dz}{\frac{dz}{2\pi i}}
\nc{\bin}[2]{\left (\begin{array}{c} {#1}\\ {#2} \end{array}\right )}
\nc{\ben}{\begin{equation}}
\nc{\een}{\end{equation}}
\nc{\bea}{\begin{eqnarray}}
\nc{\eea}{\end{eqnarray}}
\nc{\bra}[1]{\langle {#1}|}
\nc{\ket}[1]{|{#1}\rangle}
\newcommand{\Z}{\mbox{$Z\hspace{-2mm}Z$}}
\nc{\C}{\mbox{\hspace{1.24mm}\rule{0.2mm}{2.5mm}\hspace{-2.7mm} C}}
\nc{\Nat}{\mbox{\hspace{.04mm}\rule{0.2mm}{2.8mm}\hspace{-1.5mm} N}}
\newcommand{\R}{\mbox{\hspace{.04mm}\rule{0.2mm}{2.8mm}\hspace{-1.5mm} R}}

\nc{\NP}[1]{Nucl.\ Phys.\ {\bf #1}}
\nc{\PL}[1]{Phys.\ Lett.\ {\bf #1}}
\nc{\CMP}[1]{Commun.\ Math.\ Phys.\ {\bf #1}}
\nc{\PR}[1]{Phys.\ Rev.\ {\bf #1}}
\nc{\PRL}[1]{Phys.\ Rev.\ Lett.\ {\bf #1}}
\nc{\PTP}[1]{Prog.\ Theor.\ Phys.\ {\bf #1}}
\nc{\PTPS}[1]{Prog.\ Theor.\ Phys.\ Suppl.\ {\bf #1}}
\nc{\MPL}[1]{Mod.\ Phys.\ Lett.\ {\bf #1}}
\nc{\IJMP}[1]{Int.\ Jour.\ Mod.\ Phys.\ {\bf #1}}
\nc{\IM}[1]{Invent.\ Math.\ {\bf #1}}
\nc{\SJNP}[1]{Sov. J. Nucl. Phys.\ {\bf #1}}
\nc{\JHEP}[1]{J.\ High\ Energy Phys.\ {\bf #1}}

\def\vvdots{\mathinner{\mkern1mu\raise1pt\vbox{\kern7pt\hbox{.}}\mkern2mu
 \raise4pt\hbox{.}\mkern2mu\raise7pt\hbox{.}\mkern1mu}}

\def\max{{\rm max}}
\def\min{{\rm min}}

\begin{document}

\topmargin -5mm
\oddsidemargin 5mm

\begin{titlepage}
\setcounter{page}{0}

\vspace{8mm}
\begin{center}
{\huge Fusion multiplicities as polytope volumes:}
\\[.4cm]
{\huge ${\cal N}$-point and higher-genus $su(2)$ fusion}

\vspace{15mm}
{\large J{\o}rgen Rasmussen}\footnote{rasmussj@cs.uleth.ca; supported in part
by a PIMS Postdoctoral Fellowship and by NSERC} and 
{\large Mark A. Walton}\footnote{walton@uleth.ca; supported in part by NSERC}
\\[.2cm]
{\em Physics Department, University of Lethbridge,
Lethbridge, Alberta, Canada T1K 3M4}

\end{center}

\vspace{8mm}
\centerline{{\bf{Abstract}}}
\noindent
We present the first polytope volume formulas for the multiplicities
of affine fusion, the fusion in Wess-Zumino-Witten conformal field theories, 
for example. Thus, we characterise fusion multiplicities
as discretised volumes of certain convex polytopes, and write them
explicitly as multiple sums measuring those volumes.
We focus on $su(2)$, but discuss higher-point (${\cal N}>3$) and
higher-genus fusion in a general way. The method follows that of our previous
work on tensor product multiplicities, and so is based on the concepts
of generalised Berenstein-Zelevinsky diagrams, and virtual couplings.
As a by-product, we also determine necessary and sufficient
conditions for non-vanishing higher-point fusion multiplicities.
In the limit of large level, these inequalities reduce to very simple
non-vanishing conditions for the corresponding tensor product
multiplicities. Finally, we find the minimum level at which the higher-point
fusion and tensor product multiplicities coincide. 
\end{titlepage}
\newpage
\renewcommand{\thefootnote}{\arabic{footnote}}
\setcounter{footnote}{0}

\section{Introduction}

In a recent paper \cite{RW2} we have shown how 
a higher-point $su(r+1)$ tensor product
multiplicity may be expressed as a multiple sum measuring the
discretised volume of a certain convex polytope. That work 
is an extension of our previous work \cite{RW1} on ordinary three-point 
couplings where three highest weight modules are coupled to the singlet.
The number of times the singlet occurs in the decomposition is the
associated multiplicity. 
Both of these papers are based on generalisations of the famous
Berenstein-Zelevinsky (BZ) triangles \cite{BZ}. They also rely on the use
of so-called virtual couplings, that relate different (true) couplings
associated to the same tensor product.

Our long-term objective is to extend
these results to affine $su(r+1)$ fusions. Here we make a start by 
considering $su(2)$. It turns out that all our results
on ${\cal N}$-point tensor products \cite{RW2} have analogous
and level-dependent counterparts in ${\cal N}$-point fusions. 
Firstly, a fusion multiplicity admits a polyhedral combinatorial
expression, where it is characterised by the discretised volume
of a convex polytope. Secondly, this volume may be measured explicitly
expressing the fusion multiplicity as a multiple sum.

We also work out very simple, easily remembered conditions determining when
an ${\cal N}$-point $su(2)$ fusion exists, i.e., when the associated 
multiplicity is non-vanishing. For infinite level, these ``mnemo-friendly''
conditions reduce 
to even simpler ones, solving the analogous problem for tensor products.

The second part of the present work deals with the extension of the above 
results to {\em higher-genus} $su(2)$ fusions.
The first result is a characterisation of a general
genus-$h$ ${\cal N}$-point fusion multiplicity as the discretised volume of a
convex polytope. The volume is measured explicitly, whereby the fusion
multiplicitly is expressed as a multiple sum.
In order to reduce the number of summations,
we then modify our approach slightly. 
The main building blocks in these considerations are the genus-one two-point
couplings. Combining these allows one to describe
general higher-genus ${\cal N}$-point fusion multiplicities using fewer
parameters than inherent in our polytope description.
In terms of this reduced set of parameters, we provide explicit
multiple sum formulas for the generic genus zero-, one- and two-point
fusion multiplicities. 

Our expressions make manifest that the various
fusion multiplicities are non-negative integers, and are non-decreasing 
functions of the affine level.

\section{$su(2)$ ${\cal N}$-point fusion multiplicities}

Let $M_\la$ denote an integrable highest weight module of an untwisted affine
Lie algebra. The affine highest weight is uniquely specified by the 
highest weight $\la$ of the simple horizontal subalgebra
(the underlying Lie algebra), and the affine
level $k$. Fusion of two such modules may be written as
\ben
 M_\la\times M_\mu=\sum_\nu\ N_{\la,\mu}^{(k)\ \nu} M_\nu\ \ ,
\label{MM}
\een
where $N_{\la,\mu}^{(k)\ \nu}$ is the fusion multiplicity.
Determining these multiplicities is equivalent to studying the more symmetric
problem of determining the multiplicity of the singlet 
in the expansion of the triple fusion
\ben
 M_\la\times M_\mu\times M_\nu\supset N_{\la,\mu,\nu}^{(k)}M_0\ .
\label{MMM}
\een
If $\nu^+$ denotes the highest weight conjugate to 
$\nu$, we have $N_{\lambda,\mu,\nu}^{(k)} = N_{\lambda,\mu}^{(k)\ \nu^+}$.  

The associated and level-independent tensor product multiplicity 
is denoted $T_{\la,\mu,\nu}$. It is related to the fusion multiplicity as
\ben
 T_{\la,\mu,\nu}=\lim_{k\rightarrow\infty}N_{\la,\mu,\nu}^{(k)}\ .
\een

All of this extends readily to ${\cal N}$-point couplings:
\ben
  M_{\la^{(1)}}\times...\times M_{\la^{({\cal N})}}
   \supset N_{\la^{(1)},...,\la^{({\cal N})}}^{(k)} M_0\ \ ,
\een
which are the subject of the present work. In particular, we have the relation
\ben
 T_{\la^{(1)},...,\la^{({\cal N})}}=\lim_{k\rightarrow\infty}
  N_{\la^{(1)},...,\la^{({\cal N})}}^{(k)}\ .
\label{NT}
\een
In the following we will focus on $su(2)$.

For $su(2)$ the three-point fusion multiplicity is 
\ben
 N_{\la,\mu,\nu}^{(k)}=\left\{\begin{array}{ll}
  1\ \ {\rm if}\ \ 0\leq S-\la_1,\ S-\mu_1,\ 
  S-\nu_1,\ k-S\ ,\ \ \ S\equiv\hf(\la_1+\mu_1+\nu_1)\in\Z_\geq\\
 0\ \ {\rm otherwise}\end{array}\right.
\label{three}
\een
$\la_1$ denotes the finite or first Dynkin label of the weight $\la$. 
The level-independent
information (\ref{three}) is encoded in the trivial BZ triangle
\ben
\mbox{
\begin{picture}(80,40)(-10,0)
\unitlength=0.5cm
\thicklines
\put(0,0){$c$}
\put(3,0){$a$}
\put(1.5,1.5){$b$}
\end{picture}
}
\label{BZ}
\een
where 
\ben
 a=\hf(-\la_1+\mu_1+\nu_1)\in\Z_\geq\ ,\ \ \ \ b=\hf(\la_1-\mu_1+\nu_1)
   \in\Z_\geq\ ,\ \ \ \ 
  c=\hf(\la_1+\mu_1-\nu_1)\in\Z_\geq\ ,
\label{ent}
\een
and hence
\ben
 \la_1=b+c\ ,\ \ \ \ \mu_1=c+a\ ,\ \ \ \ \nu_1=a+b\ .
\label{abc}
\een
The level dependence is contained in the affine condition
\ben
 k\geq a+b+c\ .
\label{supp}
\een

In Ref. \cite{RW2} we outlined a general method for computing 
higher-point tensor product multiplicities. It is based on gluing BZ triangles
(\ref{BZ})
together using ``gluing roots'' (we refer to Ref. \cite{RW2} for details). 
An illustration is provided by the following ${\cal N}$-point diagram 
(in this example ${\cal N}$ is assumed odd):
\ben
\mbox{
\begin{picture}(100,110)(90,-25)
\unitlength=1cm
\thicklines
 \put(0,0){\line(1,2){1}}
 \put(0,0){\line(1,0){2}}
 \put(2,0){\line(-1,2){1}}
\thinlines
\put(1,0.6){\line(0,-1){1}}
\put(1,0.6){\line(2,1){1.6}}
\put(1,0.6){\line(-2,1){1}}
\thicklines
\put(-0.6,1.3){$\la^{({\cal N})}$}
\put(0.5,-0.9){$\la^{({\cal N}-1)}$}
 \put(1.47,2){
\begin{picture}(50,50)
 \put(0,0){\line(1,-2){1}}
 \put(0,0){\line(1,0){2}}
 \put(2,0){\line(-1,-2){1}}
\thinlines
\put(1,-0.6){\line(0,1){1}}
\thicklines
\put(0.5,0.6){$\la^{({\cal N}-2)}$}
\end{picture}}
\put(3.2,0){\begin{picture}(50,50)
 \put(0,0){\line(1,2){1}}
 \put(0,0){\line(1,0){2}}
 \put(2,0){\line(-1,2){1}}
\thinlines
\put(1,0.6){\line(0,-1){1}}
\put(1,0.6){\line(2,1){1}}
\put(1,0.6){\line(-2,1){1.6}}
\thicklines
\put(2.67,1.1){$\dots$}
\put(0.5,-0.9){$\la^{({\cal N}-3)}$}
\end{picture}}
\put(7,0){\begin{picture}(50,50)
 \put(0,0){\line(1,2){1}}
 \put(0,0){\line(1,0){2}}
 \put(2,0){\line(-1,2){1}}
\thinlines
\put(1,0.6){\line(0,-1){1}}
\put(1,0.6){\line(2,1){1}}
\put(1,0.6){\line(-2,1){1}}
\thicklines
\put(0.8,-0.9){$\la^{(2)}$}
\put(2.1,1.2){$\la^{(1)}$}
\end{picture}}
\end{picture}
}
\label{lll}
\een
The role of the gluing is to take care of the summation over internal weights
in a tractable way.
The dual picture of ordinary (Feynman tree-) graphs is shown in thinner lines.
Along a gluing, the opposite weights must be identified
(for higher rank $su(r+1)$ one must identify a weight with the 
{\em conjugate} weight to the opposite one, cf. \cite{RW2}). 
The weights are simply given by sums of two entries (\ref{abc}). 
Our starting point \cite{RW2} was to relax the constraint that the entries
(\ref{ent}) should be {\em non-negative} integers. A diagram of that kind
is called a generalised diagram. Any such generalised diagram, respecting the 
gluing constraints and the outer weight constraints (\ref{lll}),
will suffice as an initial diagram.
All other diagrams (associated to the same outer weights)
may then be obtained by adding integer linear combinations of so-called
virtual diagrams: adding a basis virtual diagram changes the weight
of a given internal weight by two, leaving all other internal weights and 
all outer weights unchanged. Thus, the basis virtual diagram associated
to a particular gluing is of the form:
\ben
\mbox{
\begin{picture}(100,65)(-45,-25)
\unitlength=1cm
\thicklines
\put(-2.8,0){${\cal G}\ \ \ =$}
\put(0,0){$1$}
\put(2.4,0){$1$}
 \put(0.8,0.5){$\vvdots$}
\put(1.8,0.9){$1$}
\put(2.7,0.9){$-1$}
\put(-0.9,-0.9){$-1$}
\put(0.6,-0.9){$1$}
 \put(1.5,-0.6){$\vvdots$}
\end{picture}
}
\label{gl}
\een
Enumerating the gluing roots (\ref{gl}) in (\ref{lll}) from right to left,
the associated integer coefficients in the
linear combinations are $-g_1$,...,$-g_{{\cal N}-3}$\footnote{We are using a 
slightly different notation for these variables
than that employed in \cite{RW2}.}.
Now, re-imposing the condition that all entries must be {\em non-negative}
integers, results in a set of inequalities in the entries defining
a convex polytope in the euclidean space $\R^{{\cal N}-3}$:
\bea
 0&\leq&g_1,\ \la^{(2)}_1-g_1,\ \la^{(1)}_1-g_1\ ,\nn
 0&\leq&g_2-g_1,\ \la^{(3)}_1-g_2+g_1,\ \la^{(1)}_1+\la^{(2)}_1-g_2-g_1\ ,\nn
 &\vdots&\nn
 0&\leq&g_{{\cal N}-3}-g_{{\cal N}-4},\ \la^{({\cal N}-2)}-g_{{\cal N}-3}
  +g_{{\cal N}-4},\ \la^{(1)}_1+...+\la^{({\cal N}-3)}-g_{{\cal N}-3}
  -g_{{\cal N}-4}\ ,\nn
 0&\leq&S-\la^{({\cal N}-1)}_1-g_{{\cal N}-3},\ 
  S-\la^{({\cal N})}_1-g_{{\cal N}-3},\ -S+\la^{({\cal N}-1)}_1
   +\la^{({\cal N})}_1+g_{{\cal N}-3}\ .
\label{ten}
\eea
By construction, its discretised volume is the tensor product 
multiplicity $T_{\la^{(1)},...,\la^{({\cal N})}}$.
In (\ref{ten}) we have introduced the quantity
\ben
 S\equiv\hf\sum_{l=1}^{{\cal N}}\la^{(l)}_1\in\Z_\geq\ .
\label{S}
\een
That $S$ is an integer is a consistency condition, i.e., for $S$ a
half-integer the multiplicity vanishes.

The extension to fusion is provided by supplementing the set of
inequalities (\ref{ten}) with the associated affine conditions
(cf. (\ref{supp})), one for each triangle, i.e., one for each line
in (\ref{ten}). This results in 
the following definition of a convex polytope in the euclidean space 
$\R^{{\cal N}-3}$ (the affine conditions are written on separate lines):
\bea
 0&\leq&g_1,\ \la^{(2)}_1-g_1,\ \la^{(1)}_1-g_1,\nn
 && k-\la^{(1)}_1-\la^{(2)}_1+g_1\ ,\nn
 0&\leq&g_2-g_1,\ \la^{(3)}_1-g_2+g_1,\ \la^{(1)}_1+\la^{(2)}_1-g_2-g_1,\nn
 && k-\la^{(1)}_1-...-\la^{(3)}_1+g_2+g_1\ ,\nn
 &\vdots&\nn
 0&\leq&g_{{\cal N}-3}-g_{{\cal N}-4},\ \la^{({\cal N}-2)}_1-g_{{\cal N}-3}
  +g_{{\cal N}-4},\ \la^{(1)}_1+...+\la^{({\cal N}-3)}_1-g_{{\cal N}-3}
  -g_{{\cal N}-4}, \nn
 &&k-\la^{(1)}_1-...-\la^{({\cal N}-2)}_1+g_{{\cal N}-3}+g_{{\cal N}-4}\ ,\nn
 0&\leq&S-\la^{({\cal N}-1)}_1-g_{{\cal N}-3},\ 
  S-\la^{({\cal N})}_1-g_{{\cal N}-3},\ -S+\la^{({\cal N}-1)}_1
   +\la^{({\cal N})}_1+g_{{\cal N}-3},\nn
 &&k-S+g_{{\cal N}-3}\ .
\label{fus}
\eea
By construction, its discretised volume is the associated ${\cal N}$-point
fusion multiplicity $N_{\la^{(1)},...,\la^{({\cal N})}}^{(k)}$. 
This characterisation of the fusion multiplicity is a new result.

It is stressed that (\ref{fus}) (and also (\ref{ten})) is non-unique
as it reflects our choice of initial diagram when deriving (\ref{ten}),
cf. \cite{RW1,RW2}. Any choice will define a convex polytope of the same
shape and hence discretised volume, however. Changing the initial
triangle merely corresponds to shifting the origin, or translating the 
entire polytope.

We have seen that the fusion polytope (\ref{fus}) corresponds to
``slicing out'' a convex polytope embedded in the tensor product
polytope (\ref{ten}). Thus, our approach offers a geometrical illustration
of the statement that fusion is a truncated tensor product.

The discretised volume of the convex polytope (\ref{fus}) may be measured
explicitly. In order to avoid discussing intersections of faces
we have to choose an ``appropriate order'' of summation 
(see Ref. \cite{RW1,RW2}). However, such an order is easily found.
In the following multiple sum formula we have made a straightforward 
choice: 
\bea
 N_{\la^{(1)},...,\la^{({\cal N})}}^{(k)}&=&\sum_{g_{{\cal N}-3}=
  \max\{S-\la^{({\cal N}-1)}_1-\la^{({\cal N})}_1,\ 
   -k+S\}}^{\min\{S-\la^{({\cal N}-1)}_1,\ 
   S-\la^{({\cal N})}_1\}}\nn
 &&\times\sum_{g_{{\cal N}-4}=\max\{-\la^{({\cal N}-2)}_1+g_{{\cal N}-3},\ 
    -k+\la^{(1)}_1+...+\la^{({\cal N}-2)}_1-g_{{\cal N}-3}\}}^{\min\{
  g_{{\cal N}-3},\ \la^{(1)}_1+...+\la^{({\cal N}-3)}_1-g_{{\cal N}-3}\}}\nn
 &&\vdots\nn
 &&\times\sum_{g_2=\max\{-\la^{(4)}_1+g_3,\ -k+\la^{(1)}_1+...+\la^{(4)}_1-g_3
   \}}^{\min\{g_3,\ \la^{(1)}_1+...+\la^{(3)}_1-g_3\}}\nn
 &&\times \sum_{g_1=\max\{0,\ -k+\la^{(1)}_1+\la^{(2)}_1,\ -\la^{(3)}_1+g_2,
   \ -k+\la^{(1)}_1+...+\la^{(3)}_1-g_2\}}^{\min\{\la^{(1)}_1,\ \la^{(2)}_1,
  \ g_2,\ \la^{(1)}_1+\la^{(2)}_1-g_2\}}1\ .
\label{sum}
\eea

\subsection{Conditions for non-vanishing fusion 
and tensor product multiplicities}

Here we shall present necessary and sufficient conditions determining
when an ${\cal N}$-point fusion multiplicity is non-vanishing, 
${\cal N}\geq2$.
A similar result for the associated tensor product multiplicity is
easily read off. Both sets of conditions are given as inequalities in the
(finite) Dynkin labels. The conditions for fusion depend on the level $k$.

A fusion multiplicity is non-vanishing if and only if the associated
convex polytope has a non-vanishing discretised volume. In particular,
the multiplicity is one when the polytope is a point. An analysis
of the polytope (\ref{fus}), or equivalently of the multiple sum
formula (\ref{sum}), leads to the following necessary and sufficient
conditions for the fusion multiplicity to be non-vanishing:
\bea
  0&\leq&\la^{(l)}_1,\ S-\la^{(l)}_1,\ k-\la^{(l)}_1
    \ ,\ \ \ \ \ \ \ l=1,...,{\cal N}\ ,\nn
 0&\leq&dk-S+\la^{(l_1)}_1+...+\la^{(l_{{\cal N}-2d-1})}_1\ ,
  \ \ \ \ \ \ l_m<l_n\ \ {\rm for} \ \ m<n;\ 
  1\leq d\leq\left[\frac{{\cal N}-1}{2}\right]\ .
\label{nonvF}
\eea
$[x]$ denotes the integer value of $x$, i.e., the greatest integer
less than or equal to $x$. Note that for $d=0$ the associated
inequalities reduce to $0\leq S-\la^{(l)}_1$. These latter inequalities 
have been written separately for clarity.
The upper bound on $d$ is included to avoid redundancies.

The conditions (\ref{nonvF}) may be proved by induction.
In the set of inequalities (\ref{fus}) (or equivalently in the
multiple sum formula (\ref{sum})), one eliminates one after the other
the variables $g_1,...,g_{{\cal N}-3}$. The inequalities involving
$g_1$ and $g_{{\cal N}-3}$ are different in form from those for the remaining
${\cal N}-5$ variables (\ref{fus}). 
Thus, the induction concerns the elimination of
the middle ${\cal N}-5$ variables, $g_2,...,g_{{\cal N}-4}$.
First we eliminate $g_1$, then $g_2$ etc. After having eliminated
the first $n-1$ variables, $2\leq n-1\leq {\cal N}-4$,
we have obtained the following set of inequalities
\bea
 &&0\ \leq\ \la^{(l)}_1,\ k-\la^{(l)}_1\ \ \ {\rm for}\ \ l\leq n\ ,\nn
 &&\max\{-\la^{(n+1)}_1+g_n,\ -k+\la^{(1)}_1+...+\la^{(n+1)}_1-g_n,\ 
   0,\ \hf(-k+\la^{(1)}_1+...+\la^{(n)}_1),\nn
 &&\ \ \ \ \  -dk+\la^{(l_1)}_1+...+\la^{(l_{2d})}_1\}\ \ \ {\rm for} \ \ 
   l_m<l_{m'}\leq n\ {\rm for}\ m<m';\ 2d\leq n
 \nn
 &&\ \leq\ \min\{g_n,\ \la^{(1)}_1+...+\la^{(n)}_1-g_n,\ \hf(\la^{(1)}_1+...+
  \la^{(n)}_1),\ \la^{(1)}_1+...+\la^{(n)}_1-\la^{(l)}_1,\nn
 &&\ \ \ \ \  Dk+\la^{(l_1)}_1+...+\la^{(l_{n-2D-1})}_1\}\ \ \ {\rm for} \ \ 
   l\leq n;\ l_m<l_{m'}\leq n\ {\rm for}\ m<m';\ 2D\leq n-1\ ,\ \ \ \ \ {}
\label{ind}
\eea
in addition to the original inequalities (\ref{fus}) involving
only $g_n,...,g_{{\cal N}-3}$. It is when proving (\ref{ind})
that we use induction in $n$, and we conclude that it is true for
$2\leq n-1\leq {\cal N}-4$. Eliminating the final variable
$g_{{\cal N}-3}$ results in the asserted conditions (\ref{nonvF}),
which we believe are new.

For high level $k$ a fusion reduces to a tensor product
(\ref{NT}). Necessary and sufficient conditions for a non-vanishing tensor
product multiplicity are therefore easily read off (\ref{nonvF}):
\ben
 0\leq\la^{(l)}_1,\ S-\la^{(l)}_1\ ,\ \ \ \ \ \ \ l=1,...,{\cal N}\ .
\label{nonvT}
\een
As discussed in Ref. \cite{RW2}, this result is easily verified for
${\cal N}\leq4$. For general ${\cal N}$ it is believed to be a new result.

We note that (\ref{nonvF}) and (\ref{nonvT}) are also valid for 
${\cal N}=2$, despite the fact
that, a priori, the inequalities were derived for ${\cal N}\geq3$ only.

\subsection{Conditions on the level}

The lower bound on $k$ is immediately read off (\ref{nonvF}).
In ordinary three-point fusion the analogous bound is sometimes referred
to as the minimum threshold level, and is denoted $t^\min$. It specifies
the minimum value of $k$ for which $N_{\la^{(1)},...,\la^{({\cal N})}}^{(k)}$ 
is non-vanishing:
\ben
 N_{\la^{(1)},...,\la^{({\cal N})}}^{(k<t^\min)}=0,\ \ \ \ 
 N_{\la^{(1)},...,\la^{({\cal N})}}^{(k\geq t^\min)}>0\ .
\een
It does not make sense to assign a minimum threshold level to a fusion for
which the associated tensor product multiplicity 
$T_{\la^{(1)},...,\la^{({\cal N})}}$ vanishes.

According to (\ref{nonvF}) we have
\ben
 t^\min=\max\{\la^{(l)}_1,\ \frac{1}{d}(\la^{(l_{{\cal N}-2d})}_1+...+
  \la^{(l_{{\cal N}})}_1-S)\}\ ,
\een
with the parameters specified as in (\ref{nonvF}).

The maximum threshold level, denoted $t^\max$, is defined as the minimum
level $k$ for which the fusion multiplicity equals the tensor product
multiplicity:
\ben
 N_{\la^{(1)},...,\la^{({\cal N})}}^{(k<t^\max)}<
  T_{\la^{(1)},...,\la^{({\cal N})}},\ \ \ \ 
 N_{\la^{(1)},...,\la^{({\cal N})}}^{(k\geq t^\max)}=
  T_{\la^{(1)},...,\la^{({\cal N})}}\ .
\label{tmax}
\een
Again, it is not natural to assign a maximum threshold level to a fusion if
$T_{\la^{(1)},...,\la^{({\cal N})}}$ vanishes. Though in this case, one could
define it as $t^\max=0$, since by assumption $k\in\Z_\geq$, and
(\ref{tmax}) would still be respected.

To compute $t^\max$ in our case, we first observe that all affine conditions
in (\ref{fus}) are redundant when $k\geq S$. As an illustration, we have
(assuming $3\leq m\leq {\cal N}-3$)
\bea
 0&\leq&(g_1)+(g_2-g_1)+...+(g_{m-2}-g_{m-3})\nn
  &&+(\la^{(m+1)}_1-g_m+g_{m-1})+...+(\la^{({\cal N}-2)}_1-g_{{\cal N}-3}
    +g_{{\cal N}-4})\nn
  &&+(-S+\la^{({\cal N}-1)}_1+\la^{({\cal N})}_1+g_{{\cal N}-3})\nn
 &=&\la^{(m+1)}_1+...+\la^{({\cal N})}_1-S+g_{m-1}+g_{m-2}\nn
 &\leq&k-\la^{(1)}_1-...-\la^{(m)}_1+g_{m-1}+g_{m-2}\ .
\eea
This means that $t^\max\leq S$. In order to show that
\ben
 t^\max=S\ ,
\label{tS}
\een
we first assume that there exists integer $n$, $2\leq n\leq {\cal N}-2$,
($n=1$ is trivial) such that
\ben
 \la^{(1)}_1+...+\la^{(n-1)}_1-S<0\leq\la^{(1)}_1+...+\la^{(n)}_1-S\ .
\label{n}
\een
We then consider the point defined by
\bea
 &&g_1=...=g_{n-2}=0,\nn
 &&g_{n-1}=\la^{(1)}_1+...+\la^{(n)}_1-S
   =S-\la^{(n+1)}_1-...-\la^{({\cal N})}_1,\nn
 &\vdots&\nn
 &&g_{{\cal N}-3}=\la^{(1)}_1+...+\la^{({\cal N}-2)}_1-S
    =S-\la^{({\cal N}-1)}_1-\la^{({\cal N})}_1\ .
\eea
It is straightforward to show that it is in the fusion polytope (\ref{fus})
when $k\geq S$, and that it is not when $k<S$.

Finally, if there does not exist an $n$, $2\leq n\leq {\cal N}-2$,
satisfying (\ref{n}), we must have 
$S\leq\la^{({\cal N}-1)}_1+\la^{({\cal N})}_1$. In that case we
consider the point $g_l=0,\ l=1,...,{\cal N}-3$. 
For this point to be in the polytope, the condition on $k$ (\ref{fus})
is $S\leq k$, and we conclude that the maximum threshold level is 
given by (\ref{tS}).

\section{Higher-genus $su(2)$ fusion multiplicities}

Here we will discuss the extension of our results above on genus-zero fusion
to generic genus-$h$ fusion. 
$N_{\la^{(1)},...,\la^{({\cal N})}}^{(k,h)}$ denotes the genus-$h$
${\cal N}$-point fusion multiplicity. 

Just as in the case of vanishing genus, we may choose the channel freely.
A simple extension of (\ref{lll}) is the following
genus-$h$ ${\cal N}$-point diagram (in this example ${\cal N}$ is assumed
even, while $h$ is arbitrary):
\ben
\mbox{
\begin{picture}(180,80)(105,-20)
\unitlength=0.5cm
\thicklines
 \put(0,0){\line(1,2){1}}
 \put(0,0){\line(1,0){2}}
 \put(2,0){\line(-1,2){1}}
\thinlines
\put(1,0.6){\line(0,-1){1}}
\put(1,0.6){\line(2,1){1.6}}
\put(1,0.6){\line(-2,1){1}}
\thicklines
\put(-1.1,1.2){$\la^{({\cal N})}$}
\put(0.1,-1.3){$\la^{({\cal N}-1)}$}
 \put(1.34,2){
\begin{picture}(50,50)
 \put(0,0){\line(1,-2){1}}
 \put(0,0){\line(1,0){2}}
 \put(2,0){\line(-1,-2){1}}
\thinlines
\put(1,-0.6){\line(0,1){1}}
\thicklines
\put(0.1,0.6){$\la^{({\cal N}-2)}$}
\end{picture}}
\put(3.2,0){\begin{picture}(50,50)
 \put(0,0){\line(1,2){1}}
 \put(0,0){\line(1,0){2}}
 \put(2,0){\line(-1,2){1}}
\thinlines
\put(1,0.6){\line(0,-1){1}}
\put(1,0.6){\line(2,1){1}}
\put(1,0.6){\line(-2,1){1.6}}
\thicklines
\put(2.45,1.1){$\dots$}
\put(0.1,-1.3){$\la^{({\cal N}-3)}$}
\end{picture}}
\put(7,0){\begin{picture}(50,50)
 \put(0,0){\line(1,2){1}}
 \put(0,0){\line(1,0){2}}
 \put(2,0){\line(-1,2){1}}
\thinlines
\put(1,0.6){\line(0,-1){1}}
\put(1,0.6){\line(2,1){0.8}}
\put(1,0.6){\line(-2,1){1}}
\thicklines
\put(0.6,-1.3){$\la^{(1)}$}
\put(2.2,-0.1){0}
\put(1.2,1.9){0}
\end{picture}}
\put(10.5,1){\begin{picture}(100,80)
\unitlength=0.9cm
\put(0,0.5){\line(0,-1){1}}
\put(0,0.5){\line(2,-1){1}}
\put(0,-0.5){\line(2,1){1}}
\put(1.5,0){\line(2,1){1}}
\put(1.5,0){\line(2,-1){1}}
\put(2.5,0.5){\line(0,-1){1}}
\thinlines
\put(1.25,0){\circle{1.4}}
\put(0.55,0){\line(-1,0){1.5}}
\put(1.95,0){\line(1,0){1.05}}
\thicklines
 \put(3.5,0){$\dots$}
\put(5,0.5){\line(0,-1){1}}
\put(5,0.5){\line(2,-1){1}}
\put(5,-0.5){\line(2,1){1}}
\put(6.5,0){\line(2,1){1}}
\put(6.5,0){\line(2,-1){1}}
\put(7.5,0.5){\line(0,-1){1}}
\thinlines
\put(6.25,0){\circle{1.4}}
\put(5.55,0){\line(-1,0){1.05}}
\put(6.95,0){\line(1,0){1.05}}
\thicklines
\put(9.15,0){\line(-2,1){1}}
\put(9.15,0){\line(-2,-1){1}}
\put(8.15,0.5){\line(0,-1){1}}
\thinlines
\put(9.15,0){\circle{1.4}}
\put(8,0){\line(1,0){0.45}}
\end{picture}}
\end{picture}
}
\label{gN}
\een
Again, the dual trivalent fusion graph is represented by thinner lines 
and loops. $h$ is the number of such loops or handles. 
The role of the two zeros in (\ref{gN}) will be discussed below.

Independent of the choice of channel, the number of internal weights
or gluings is ${\cal N}+3(h-1)$, while the number of vertices or triangles
is ${\cal N}+2(h-1)$.

The basis diagram associated to the ``self-coupling'' or tadpole diagram
\ben
\mbox{
\begin{picture}(100,40)(-25,-20)
\unitlength=0.7cm
\thicklines
\put(1.15,0){\line(-2,1){1}}
\put(1.15,0){\line(-2,-1){1}}
\put(0.15,0.5){\line(0,-1){1}}
\thinlines
\put(1.15,0){\circle{1.4}}
\put(-0.5,0){\line(1,0){0.95}}
\end{picture}
}
\label{tad}
\een
is
\ben
\mbox{
\begin{picture}(100,40)(-25,-20)
\unitlength=0.7cm
\thicklines
\put(0,0.5){0}
\put(0,-0.5){0}
\put(1,0){1}
\end{picture}
}
\label{gl2}
\een
We call (\ref{gl2}) a loop-gluing diagram.
It is stressed that it differs from the gluing root (\ref{gl})
since it adds only {\em one} to the internal weight and not two.
This discrepancy follows from the fact that the Dynkin labels satisfy
$\la_1+\mu_1+\nu_1\in2\Z_\geq$, so if two weights are changed simultaneously
and equally, we can only require an even change of the {\em sum} of them.

A similar situation arises when considering the genus-one two-point
coupling
\ben
\mbox{
\begin{picture}(100,50)(0,-20)
\unitlength=1cm
\thicklines
\put(0,0.5){\line(0,-1){1}}
\put(0,0.5){\line(2,-1){1}}
\put(0,-0.5){\line(2,1){1}}
\put(1.5,0){\line(2,1){1}}
\put(1.5,0){\line(2,-1){1}}
\put(2.5,0.5){\line(0,-1){1}}
\thinlines
\put(1.25,0){\circle{1.4}}
\put(0.55,0){\line(-1,0){1.05}}
\put(1.95,0){\line(1,0){1.05}}
\thicklines
\put(-0.9,-0.1){$\la$}
\put(3.2,-0.1){$\mu$}
\end{picture}
}
\label{two}
\een
A simple analysis shows that there are two basis loop-gluings associated
to this coupling, and that they may be represented by the diagrams
\ben
\mbox{
\begin{picture}(100,40)(60,-20)
\unitlength=0.7cm
\thicklines
\put(-2,0){${\cal L}\ =$}
\put(0,0.5){0}
\put(0,-0.5){0}
\put(1,0){1}
\put(2,0){1}
\put(3,0.5){0}
\put(3,-0.5){0}
\put(6,0){${\cal L}'\ =$}
\put(8,0.5){1}
\put(7.9,-0.5){-1}
\put(9,0){0}
\put(10,0){0}
\put(11,0.5){1}
\put(10.9,-0.5){-1}
\end{picture}
}
\label{lg}
\een

It is now easy to write down the inequalities defining the convex polytope.
Our choice of initial diagram is indicated in (\ref{gN}) by the two zeros:
all entries of the higher-genus part to the right of them are zero,
while the ${\cal N}$-point part follows the pattern of the initial diagram
associated to (\ref{lll}) and (\ref{ten}) - see \cite{RW2} for details.
Enumerating the (loop-)gluings from right to left (and ${\cal L}$ before
${\cal L}'$), the integer coefficients in the
linear combinations are $g_1,...,g_{h},-g_{h+1},...,-g_{{\cal N}+h-2}$
(the sign convention is merely for convenience), and $l_1,l'_1,...,
l_{h-1},l_{h-1}'$, while $l$ is associated to the
tadpole at the extreme right. 
Listing the inequalities associated to the triangles from right to left,
we have the following convex polytope (assuming $h\geq1$):
\bea
 0&\leq&l-g_1,\ g_1,\ g_1,\nn
  &&k-g_1-l\ ,\nn
 0&\leq&l_1-g_1,\ g_1+l_1',\ g_1-l_1',\nn
  &&k-g_1-l_1\ ,\nn
 0&\leq&l_1-g_2,\ g_2+l_1',\ g_2-l_1',\nn
  &&k-g_2-l_1\ ,\nn
 &\vdots&\nn
 0&\leq&l_{h-1}-g_{h-1},\ g_{h-1}+l_{h-1}',\ g_{h-1}-l_{h-1}',\nn
  &&k-g_{h-1}-l_{h-1}\ ,\nn
 0&\leq&l_{h-1}-g_h,\ g_h+l_{h-1}',\ g_h-l_{h-1}',\nn
  &&k-g_h-l_{h-1}\ ,\nn
 0&\leq&g_{h+1}+g_{h},\ -g_{h+1}+g_{h},\ \la^{(1)}_1-g_{h+1}-g_{h},\nn
  &&k-\la^{(1)}_1+g_{h+1}-g_{h}\ ,\nn
 0&\leq&g_{h+2}-g_{h+1},\ \la^{(1)}_1-g_{h+2}-g_{h+1},\ \la^{(2)}_1
   -g_{h+2}+g_{h+1},\nn
  &&k-\la^{(1)}_1-\la^{(2)}_1+g_{h+2}+g_{h+1}\ ,\nn
 &\vdots&\nn
 0&\leq&g_{{\cal N}+h-2}-g_{{\cal N}+h-3},\ \la^{(1)}_1+...
   +\la^{({\cal N}-3)}_1-g_{{\cal N}+h-2}-g_{{\cal N}+h-3},\nn 
  &&\la^{({\cal N}-2)}_1-g_{{\cal N}+h-2}+g_{{\cal N}+h-3},\nn
  &&k-\la^{(1)}_1-...-\la^{({\cal N}-2)}_1+g_{{\cal N}+h-2}
    +g_{{\cal N}+h-3}\ ,\nn
 0&\leq&S-\la^{({\cal N}-1)}_1-g_{{\cal N}+h-2},\ S-\la^{({\cal N})}_1
  -g_{{\cal N}+h-2},\ -S+\la^{({\cal N}-1)}_1+\la^{({\cal N})}_1
   +g_{{\cal N}+h-2},\nn
  &&k-S+g_{{\cal N}+h-2}\ .
\label{polgN}
\eea
By construction, its discretised volume is the fusion multiplicity
$N_{\la^{(1)},...,\la^{({\cal N})}}^{(k,h)}$, which then provides a
new way of characterising fusion multiplicities.
The volume may be measured explicitly expressing 
$N_{\la^{(1)},...,\la^{({\cal N})}}^{(k,h)}$ as a multiple sum:
\bea
 N_{\la^{(1)},...,\la^{({\cal N})}}^{(k,h)}&=&\sum_{g_{{\cal N}+h-2}}...
  \sum_{g_h}\left(\sum_{l_{h-1}'}\sum_{l_{h-1}}\sum_{g_{h-1}}\right)...
  \left(\sum_{l_1'}\sum_{l_1}\sum_{g_1}\right)\sum_l1\ .
\label{sumNh}
\eea
The summation variables are bounded according to
\bea
 g_1\leq&l&\leq k-g_1\ ,\nn
 |l_1'|\leq &g_1&\leq \min\{l_1,\ k-l_1\}\ ,\nn
 g_2\leq &l_1&\leq k-g_2\ ,\nn
 -g_2\leq &l_1'&\leq g_2\ ,\nn
 &\vdots&\nn
 |l_{h-1}'|\leq &g_{h-1}&\leq \min\{l_{h-1},\ k-l_{h-1}\}\ ,\nn
 g_h\leq &l_{h-1}&\leq k-g_h\ ,\nn
 -g_h\leq &l_{h-1}'&\leq g_h\ ,\nn
 |g_{h+1}|\leq&g_h&\leq\min\{\la_1^{(1)}-g_{h+1},\ k-\la_1^{(1)}+g_{h+1}\}
   \ ,\nn
 \max\{-\la_1^{(2)}+g_{h+2},\ \ \ \ \ \hspace{1cm}&&\nn
   -k+\la_1^{(1)}+\la_1^{(2)}-g_{h+2}\}
  \leq&g_{h+1}&\leq \min\{g_{h+2},\ \la_1^{(1)}-g_{h+2}\}\ ,\nn
 &\vdots&\nn
 \max\{-\la_1^{({\cal N}-2)}+g_{{\cal N}+h-2},\ \ \ \hspace{2cm} &&\nn
  -k+\la_1^{(1)}+...+ 
  \la_1^{({\cal N}-2)}-g_{{\cal N}+h-2}\}\leq &g_{{\cal N}+h-3}&\leq
  \min\{g_{{\cal N}+h-2},\nn
    &&\ \ \hspace{1cm}\la_1^{(1)}+...+\la_1^{({\cal N}-3)}-
  g_{{\cal N}+h-2}\}\ ,\nn
 \max\{S-\la_1^{({\cal N}-1)}-\la_1^{({\cal N})},\ -k+S\}\leq
  &g_{{\cal N}+h-2}&\leq\min\{S-\la_1^{({\cal N}-1)},\ S-\la_1^{({\cal N})}\}
  \ .
\label{sumNhb}
\eea
This constitutes the first explicit result for the general genus-$h$ 
${\cal N}$-point fusion multiplicities. In the following we
will discuss a few examples, where the convex polytope characterisation
is sacrifised in order to reduce the number of summation variables.

\subsection{Two-point couplings}

Let us first consider the genus-one two-point coupling (\ref{two}).
According to the general discussion above, one may express the associated 
fusion multiplicity in terms of two parameters. A further analysis shows that
\ben
 N_{\la,\mu}^{(k,1)}=\left\{\begin{array}{lll}
   (\min\{\la_0,\mu_0\}+1)(\min\{\la_1,\mu_1\}+1)\ ,
    \ \ \ &|\la_1-\mu_1|\in2\Z_\geq\\
 0\ ,\ \ \ &|\la_1-\mu_1|+1\in2\Z_\geq  \end{array}\right.
\label{onetwo}
\een
where the zero'th Dynkin label of the affine weight $\la$ is $\la_0=k-\la_1$.
Now, it is straightforward to construct higher-genus two-point diagrams
by gluing together diagrams like (\ref{two}) 
\ben
\mbox{
\begin{picture}(100,50)(70,-20)
\unitlength=1cm
\thicklines
\put(-0.9,-0.1){$\la$}
\put(0,0.5){\line(0,-1){1}}
\put(0,0.5){\line(2,-1){1}}
\put(0,-0.5){\line(2,1){1}}
\put(1.5,0){\line(2,1){1}}
\put(1.5,0){\line(2,-1){1}}
\put(2.5,0.5){\line(0,-1){1}}
\thinlines
\put(1.25,0){\circle{1.4}}
\put(0.55,0){\line(-1,0){1.05}}
\put(1.95,0){\line(1,0){1.05}}
\thicklines
 \put(3.5,0){$\dots$}
\put(5,0.5){\line(0,-1){1}}
\put(5,0.5){\line(2,-1){1}}
\put(5,-0.5){\line(2,1){1}}
\put(6.5,0){\line(2,1){1}}
\put(6.5,0){\line(2,-1){1}}
\put(7.5,0.5){\line(0,-1){1}}
\thinlines
\put(6.25,0){\circle{1.4}}
\put(5.55,0){\line(-1,0){1.05}}
\put(6.95,0){\line(1,0){1.05}}
\thicklines
\put(8.2,-0.1){$\mu$}
\end{picture}
}
\label{twog}
\een
When computing the associated
fusion multiplicities one uses the result (\ref{onetwo}), paying attention
to the finite Dynkin labels being odd or even.
For example, when $\la_1$ and $\mu_1$ are both even, the sum formula reads
\bea
 N_{\la,\mu}^{(k,h)}&=&\sum_{m_1,...,m_{h-1}=0}^{[k/2]}
  (k-\max\{\la_1,2m_1\}+1) (\min\{\la_1,2m_1\}+1)\nn
 &&\times
  (k-\max\{2m_1,2m_2\}+1)(\min\{2m_1,2m_2\}+1)\nn
 &&\vdots\nn
 &&\times(k-\max\{2m_{h-2},2m_{h-1}\}+1)(\min\{2m_{h-2},2m_{h-1}\}+1)\nn
 &&\times(k-\max\{2m_{h-1},\mu_1\}+1)(\min\{2m_{h-1},\mu_1\}+1)\ . 
\label{sumtwo}
\eea
It is easily adjusted to cover the situation
when both labels are odd (see also (\ref{sumzero2})). 
If one label is odd and the 
other is even, the associated fusion multiplicity vanishes.
Note that the number of summation variables is $h-1$, while the number
of summations in our previous treatment (\ref{sumNh}) was $3h-1$.
Thus, from that point of view (\ref{sumtwo}) is a considerable simplification.

The summations in (\ref{sumtwo}) are, in principle, 
straightforward to evaluate using the formula
\ben
 \sum_{m=1}^M(m)_{s}=\frac{1}{s+1}(M)_{s+1}\ ,
\label{sumf}
\een
where 
\ben
 (a)_n\equiv a(a+1)...(a+n-1)\ .
\een
(\ref{sumf}) is easily proven by induction.

\subsection{One-point couplings}

A one-point coupling simply corresponds to putting one of the weights
of a two-point coupling equal to zero. It may be illustrated by the diagram
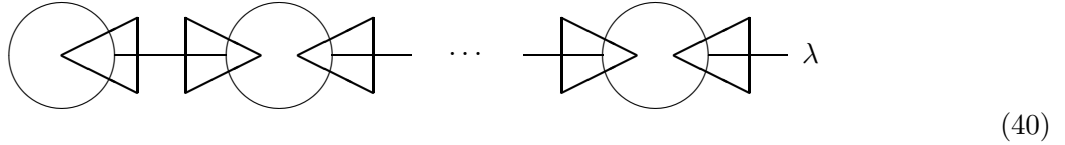
\ben
\mbox{
\begin{picture}(100,50)(100,-30)
\unitlength=1cm
\thicklines
\put(0,0){\line(2,1){1}}
\put(0,0){\line(2,-1){1}}
\put(1,0.5){\line(0,-1){1}}
\thinlines
\put(0,0){\circle{1.4}}
\put(0.7,0){\line(1,0){0.45}}
\thicklines
\end{picture}
\put(1.7,0){\begin{picture}(100,80)(155,-30)
\unitlength=1cm
\thicklines
\put(0,0.5){\line(0,-1){1}}
\put(0,0.5){\line(2,-1){1}}
\put(0,-0.5){\line(2,1){1}}
\put(1.5,0){\line(2,1){1}}
\put(1.5,0){\line(2,-1){1}}
\put(2.5,0.5){\line(0,-1){1}}
\thinlines
\put(1.25,0){\circle{1.4}}
\put(0.55,0){\line(-1,0){1.05}}
\put(1.95,0){\line(1,0){1.05}}
\thicklines
 \put(3.5,0){$\dots$}
\put(5,0.5){\line(0,-1){1}}
\put(5,0.5){\line(2,-1){1}}
\put(5,-0.5){\line(2,1){1}}
\put(6.5,0){\line(2,1){1}}
\put(6.5,0){\line(2,-1){1}}
\put(7.5,0.5){\line(0,-1){1}}
\thinlines
\put(6.25,0){\circle{1.4}}
\put(5.55,0){\line(-1,0){1.05}}
\put(6.95,0){\line(1,0){1.05}}
\thicklines
\put(8.2,-0.1){$\la$}
\end{picture}}
}
\label{oneg}
\een
and the associated fusion multiplicity is
\bea
 N_\la^{(k,h)}&=&\sum_{m_1,...,m_{h-1}=0}^{[k/2]}(k-2m_1+1)\nn
 &&\times
  (k-\max\{2m_1,2m_2\}+1)(\min\{2m_1,2m_2\}+1)\nn
 &&\vdots\nn
 &&\times(k-\max\{2m_{h-2},2m_{h-1}\}+1)(\min\{2m_{h-2},2m_{h-1}\}+1)\nn
 &&\times(k-\max\{2m_{h-1},\la_1\}+1)(\min\{2m_{h-1},\la_1\}+1)\ ,\ \ \ \ 
 \ \ \ \ \ \ \ \ \ \ \   \la_1\in2\Z_\geq\ .
\label{sumone}
\eea
It is noted that the Dynkin label $\la_1$ must be even. For $h=1$ 
(\ref{sumone}) reduces to
\ben
 N_\la^{(k,1)}=\left\{\begin{array}{lll}
  k-\la_1+1\ ,\ \ \ &\la_1\in2\Z_\geq\\
 0\ ,\ \ \ &\la_1+1\in2\Z_\geq \end{array}\right.
\een

\subsection{Zero-point couplings}

As for any other ${\cal N}$, there are many possible choices of channels when
discussing zero-point couplings. An immediate application of our
discussion on two-point couplings (\ref{twog}) corresponds to the diagram
\ben
\mbox{
\begin{picture}(100,60)(110,-30)
\unitlength=1cm
\thicklines
\put(0,0){\line(2,1){1}}
\put(0,0){\line(2,-1){1}}
\put(1,0.5){\line(0,-1){1}}
\thinlines
\put(0,0){\circle{1.4}}
\put(0.7,0){\line(1,0){0.45}}
\thicklines
\end{picture}
\put(1.7,0){\begin{picture}(100,80)(155,-30)
\unitlength=1cm
\thicklines
\put(0,0.5){\line(0,-1){1}}
\put(0,0.5){\line(2,-1){1}}
\put(0,-0.5){\line(2,1){1}}
\put(1.5,0){\line(2,1){1}}
\put(1.5,0){\line(2,-1){1}}
\put(2.5,0.5){\line(0,-1){1}}
\thinlines
\put(1.25,0){\circle{1.4}}
\put(0.55,0){\line(-1,0){1.05}}
\put(1.95,0){\line(1,0){1.05}}
\thicklines
 \put(3.5,0){$\dots$}
\put(5,0.5){\line(0,-1){1}}
\put(5,0.5){\line(2,-1){1}}
\put(5,-0.5){\line(2,1){1}}
\put(6.5,0){\line(2,1){1}}
\put(6.5,0){\line(2,-1){1}}
\put(7.5,0.5){\line(0,-1){1}}
\thinlines
\put(6.25,0){\circle{1.4}}
\put(5.55,0){\line(-1,0){1.05}}
\put(6.95,0){\line(1,0){1.05}}
\thicklines
\put(9.15,0){\line(-2,1){1}}
\put(9.15,0){\line(-2,-1){1}}
\put(8.15,0.5){\line(0,-1){1}}
\thinlines
\put(9.15,0){\circle{1.4}}
\put(8,0){\line(1,0){0.45}}
\end{picture}}
}
\label{zerog1}
\een
This is obtained by putting both weights in (\ref{twog})
equal to zero, and the associated fusion multiplicity may be expressed as
\bea
 N^{(k,h)}&=&\sum_{m_1,...,m_{h-1}=0}^{[k/2]}(k-2m_1+1)\nn
 &&\times
  (k-\max\{2m_1,2m_2\}+1)(\min\{2m_1,2m_2\}+1)\nn
 &&\vdots\nn
 &&\times(k-\max\{2m_{h-2},2m_{h-1}\}+1)(\min\{2m_{h-2},2m_{h-1}\}+1)\nn
 &&\times(k-2m_{h-1}+1)\ .
\label{sumzero1}
\eea
Another ``natural'' channel is governed by the diagram
\ben
\mbox{
\begin{picture}(100,75)(70,-20)
\unitlength=1cm
\thicklines
\put(0,0.5){\line(0,-1){1}}
\put(0,0.5){\line(2,-1){1}}
\put(0,-0.5){\line(2,1){1}}
\put(1.5,0){\line(2,1){1}}
\put(1.5,0){\line(2,-1){1}}
\put(2.5,0.5){\line(0,-1){1}}
\thinlines
\put(1.25,0){\circle{1.4}}
\put(0.55,0){\line(-1,0){1.55}}
\put(1.95,0){\line(1,0){1.05}}
\thicklines
 \put(3.5,0){$\dots$}
\put(5,0.5){\line(0,-1){1}}
\put(5,0.5){\line(2,-1){1}}
\put(5,-0.5){\line(2,1){1}}
\put(6.5,0){\line(2,1){1}}
\put(6.5,0){\line(2,-1){1}}
\put(7.5,0.5){\line(0,-1){1}}
\thinlines
\put(6.25,0){\circle{1.4}}
\put(5.55,0){\line(-1,0){1.05}}
\put(6.95,0){\line(1,0){1.55}}
 \put(3.75,0){\oval(9.5,3)[t]}
\end{picture}
}
\label{zerog2}
\een
Following our general prescription above for computing the associated
fusion multiplicity, results in the expression
\bea
 N^{(k,h)}&=&\sum_{m_1,...,m_{g-1}=0}^{[k/2]}
  (k-\max\{2m_{h-1},2m_1\}+1) (\min\{2m_{h-1},2m_1\}+1)\nn
 &&\times
  (k-\max\{2m_1,2m_2\}+1)(\min\{2m_1,2m_2\}+1)\nn
 &&\vdots\nn
 &&\times(k-\max\{2m_{h-2},2m_{h-1}\}+1)(\min\{2m_{h-2},2m_{h-1}\}+1)\nn
 &+&\sum_{m_1,...,m_{h-1}=0}^{[(k-1)/2]}
  (k-\max\{2m_{h-1},2m_1\}) (\min\{2m_{h-1},2m_1\}+2)\nn
 &&\times
  (k-\max\{2m_1,2m_2\})(\min\{2m_1,2m_2\}+2)\nn
 &&\vdots\nn
 &&\times(k-\max\{2m_{h-2},2m_{h-1}\})(\min\{2m_{h-2},2m_{h-1}\}+2)
\label{sumzero2}
\eea
which differs considerably in form from (\ref{sumzero1}).
Nevertheless, by construction, the two multiple sums must be identical.
We will not attempt to prove that explicitly. This identity provides a
simple example of the result of identifying the fusion multiplicities
computed using different channels.

Examples of non-trivial zero-point fusion multiplicities are 
\ben
 N^{(k,1)}=k+1\ ,
\een
and
\ben
 N^{(k,2)}=\frac{(k+1)_3}{6}=\frac{1}{6}(k+1)(k+2)(k+3)\ .
\een

\section{Comments}

We conclude by adding a few comments, primarily on the existing literature.
In Ref. \cite{Dow} Dowker discusses results on fusion
multiplicities based on the Verlinde formula \cite{Ver}. The results are
expressed in terms of twisted cosec sums and Bernoulli polynomials, and 
pertain essentially to two-point couplings (and therefore also to one-
and zero-point couplings).
Particular emphasis is put on the classical limit where the level
$k$ tends to infinity, and previous results on that limit
are recovered (\cite{Kil,Wit} for zero-point couplings and \cite{Wit}
for one-point couplings). In the language employed in \cite{Dow},
fusion multiplicities correspond to dimensions of certain vector bundles
over the moduli space of an ${\cal N}$-punctured Riemann surface of
genus $h$.

The results of \cite{Dow} are essentially obtained by trigonometric
manipulations of the Verlinde formula. They do not, therefore, display
any transparent relationship with our convex polytope approach.
Nevertheless, a comparison of results leads to interesting identities
between different types of multiple sums, and some similarities of
the final expressions are apparent. One could try to prove 
their equivalence by brute force. That is beyond the scope of the present
work, though.

The results of \cite{Dow} do not offer an immediate resolution to the 
question of when a fusion multiplicity is non-vanishing. By construction, a
characterisation in terms of a convex polytope, on the other hand,
is ``almost'' designed to address such problems. Furthermore, our approach
seems amenable to the treatment of higher rank $su(r+1)$ fusions, whereas 
an application of the Verlinde formula appears technically very complicated.
We are currently considering such an extention of our approach based on
previous results on the role of BZ triangles in affine $su(3)$ and $su(4)$
fusions \cite{BMW,BKMW}. A different approach to fusion based
on the depth rule and the correspondence to three-point functions in 
Wess-Zumino-Witten 
conformal field theory may be found in our recent work \cite{Ras,RW3}.

In Ref. \cite{Kir}, Kirillov provides a combinatorial formula for the 
${\cal N}$-point $su(2)$ fusion multiplicities. It is a fermionic-type 
formula, a sum of products of binomial coefficients, derived by applying  
the Bethe ansatz to certain solvable lattice models. (For a nice,  
brief review of formulas of fermionic and bosonic type, see the 
introduction to \cite{JM}.) No 
formulas for higher-genus multiplicities are given, however.  

Kirillov's fermionic formula has also been generalised somewhat. See 
Theorem 6.2 of \cite{SS} for a $q$-deformed $su(r+1)$ generalisation, 
and the extensive bibliography of \cite{HKOTT}. 
Although interesting for other reasons, these formulas are only valid for 
certain representations at the ${\cal N}$-points, 
and they are also restricted to $h=0$. Such restrictions do
not appear to be necessary in our method. 
\\[.4cm]
\noindent {\bf Acknowledgement} We thank Pierre Mathieu for comments.

\end{document}